\documentclass{raa}           

\usepackage{graphicx,times}
\usepackage{natbib}
\usepackage{amssymb,amsmath}

\begin{document}

   \title{On the Neutron Star/Black Hole Mass Gap and Black Hole Searches}

 \volnopage{ {\bf 20XX} Vol.\ {\bf X} No. {\bf XX}, 000--000}
   \setcounter{page}{1}

   \author{Yong Shao\inst{1,2}
   }

   \institute{Department of Astronomy, Nanjing University, Nanjing 210023,  
China; {\it shaoyong@nju.edu.cn}\\
        \and
             Key laboratory of Modern Astronomy and Astrophysics (Nanjing University), Ministry of
Education, Nanjing 210023, China \\
\vs \no
   {\small Received 20XX Month Day; accepted 20XX Month Day}
}

\abstract{Mass distribution of black holes in low-mass X-ray binaries previously suggested 
the existence of a $ \sim 2-5M_{\odot} $ 
mass gap between the most massive neutron stars and the least massive black holes, 
while some recent evidence appears
to support that this mass gap is being populated. Whether there is a mass gap or not can potentially 
shed light on the physics of supernova
explosions that form neutron stars and black holes, although significant mass accretion of neutron stars 
including binary mergers may lead to the formation of mass-gap objects.
In this review, I collect the compact objects that are 
probable black holes with masses
being in the gap. Most of them are in binaries, their mass measurements are 
obviously subject to some uncertainties. Current observations
are still unable to confidently infer an absence or presence of the mass gap. 
Ongoing and future surveys are expected to build the mass
spectrum of black holes which can be used to constrain the process of their formation especially in binaries.
I describe the theoretical predictions for the formation of black holes in various types of binaries, 
and present some prospects of searching for black holes via electromagnetic and gravitational wave observations. 
}

   \authorrunning{Y. Shao}            
   \titlerunning{On the NS/BH Mass Gap and BH Searches}  
   \maketitle

%
\section{Introduction}           

Stars with masses $ \gtrsim 8M_{\odot} $ are believed to undergo core collapse at the end of their lives and leave behind compact objects 
as either neutron stars (NSs) or black holes (BHs). The mass distribution of compact objects contains the imprints of the physics of 
core-collapse supernovae. Almost all NSs with mass measurements are in binaries \citep[see][for a review]{ozel2016}, and 
more than 100 such NSs have been reported to date \citep{Alsing2018,Shao2020} including those from gravitational wave 
transients \citep{lvk2021}. Recently, gravitational microlensing observations start to reveal the nature of
possible NSs with measured lens masses  \citep{Kaczmarek2022}.  
Based on electromagnetic observations,   
the maximal mass of NSs is $ \sim 2M_{\odot} $ in the binary pulsars with a white-dwarf (WD) 
companion \citep{Demorest2010,Antoniadis2013,Cromartie2020,Farr2020}
and it likely rises up to $ \sim 2.4M_{\odot} $ in the eclipsing binaries with a main-sequence 
companion \citep{Clark2002,Linares2018,Kandel2020,Romani2022}. These are agreement 
with the maximum stable NS mass inferred from gravitational wave observations of the first NS+NS merger GW170817
\citep{Margalit2017,Rezzolla2018,Ruiz2018,Shibata2019}. In gravitational wave transients except GW190814 and GW190425, 
the NS components have the masses of $ \lesssim 2.2M_{\odot} $ \citep{lvk2021,Li2021,Zhu2022}.
Over 20 BHs have been identified in X-ray binaries 
since their dynamical masses are measured \citep{Remillard2006,Casares2014}. The mass distribution of these BHs
indicates a minimal mass of $ \sim 5M_{\odot} $ \citep{Bailyn1998,ozel2010,Farr2011}. 
Also, \citet{ozel2010} demonstrated that observational selection effects are unlikely to significantly bias the observed distribution of BH masses.
It is thus suggested that there is an absence of BHs
in the mass range of $\sim 2-5M_{\odot} $, which is referred to as the mass gap between the heaviest NSs and the lightest BHs.

It is firmly established that stellar masses have a smooth distribution \citep{Salpeter1955,Kroupa1993},  the observed mass gap for compact objects 
implies a discontinuous dependence 
of the remnant masses with their progenitor masses. Some possible explanations for this gap emerge, involving a range of different supernova mechanisms 
\citep[e.g.,][]{Fryer2012,Ugliano2012,Kochanek2014,Liu2021}. Massive star observations show that most of them are born as 
members of binary and multiple systems \citep{Sana2012,Kobulnicky2014,Moe2017}. As a consequence, a significant fraction of compact objects are expected to undergo 
a binary interaction during their progenitor evolution, which complicates the understanding of massive-star evolution and
supernova-explosion processes \citep{Langer2012}. 
Binary population synthesis \citep[see][for a review]{Han2020} methods include the modelling of stellar evolution and binary interactions, 
which are widely applied to explore the population properties of a
specific type of binary systems. With this method,  
an early investigation on the mass distribution of compact remnants in
Galactic X-ray binaries showed that the rapid supernova mechanism leads to the depletion of remnants in the $ 2-5M_{\odot} $ mass range 
while the delayed mechanism results in a continuous mass distribution between NSs and BHs \citep{Belczynski2012}. 
Moreover, this gap
may appear if the vast majority of the binaries with low-mass BHs are disrupted due to a supernova-driven natal kick \citep{Mandel2021}. 
On the other hand, \citet{Kreidberg2012} suggested that 
the previously reported masses of BHs in low-mass X-ray binaries may be overestimated and actually fill the gap when considering 
possible systematic errors from the inclination measurements of the binary orbits. 
In addition, it is possible that NSs in low-mass X-ray binaries 
significantly increase their masses during
mass-transfer episodes and therefore intrude into the gap via accretion-induced collapse \citep{Gao2022}. Further precise mass 
measurements of BHs in the current sample of low-mass X-ray binaries are needed to reassess 
the existence of the mass gap and settle the mechanism of supernova explosions 
\citep[e.g.,][]{Casares2022}.

In recent years,
searching for BHs outside X-ray binaries has became a particularly active field of research. Theoretical studies predict that there are about
$ 10^{8} $ stellar-mass BHs in the Milky Way \citep[e.g.,][]{Brown1994,Timmes1996}, most of them are expected to appear as singles because of 
a binary disruption or a binary merger \citep{Wiktorowicz2019,Olejak2020}. Gravitational microlensing has been proposed to find dark compact objects 
over past decades \citep{Paczynski1986,Paczynski1996}, and already extended to various searches of stellar-mass BHs 
\citep[e.g.,][]{Mao2002,Bennett2002,Wyrzykowski2016}. With gravitational microlensing, optical surveys  have been carried out to probe a wide range 
of stellar-remnant masses  \citep{Wyrzykowski2020,mw21,Mroz2021}. In binary systems, radial velocity searches of optical
companions have been proposed to discover quiescent BHs \citep{Guseinov1966,Trimble1969,Gu2019}. 
Recently, a series of optical surveys are implemented to identify these types of BHs in the Galactic field 
\citep[e.g.,][]{Liu2019,Thompson2019,Zheng2019,Rowan2021,Gomel2022,Gaia2022,Fu2022,Jayasinghe2022,Mahy2022,Shahaf2022,Tanikawa2022}, 
globular clusters \citep{Giesers2018,Giesers2019}, and the Large Magellanic Cloud \citep{Shenar2022a,Shenar2022b}.
The discovery of the first binary BH merger GW150914 \citep{Abbott2016} opens a new window to detect BHs in gravitational waves, and until now over
90 binary mergers with at least a BH component have been reported \citep{Abbott2019,Abbott2021,lvk2021,Nitz2021}. 
For some recently identified compact objects, there is growing evidence that the mass gap is being populated. 
It is predicted that ongoing and future surveys can greatly increase the number of the compact objects with mass measurements and build a mass
spectrum to resolve the issue of mass gap.

This review is organized as follows. In Section 2, I collect the (candidate) BHs with masses in the gap from the literature  (see also Figure~1)
and briefly discuss their properties. I describe the formation scenarios of BH binaries in Section 3, and 
discuss the detection prospects for various types of Galactic BH binaries (see also Table~1) in Section 4.


\section{Observations of probable mass-gap black holes}

In Figure~1, I present the mass distribution of probable mass-gap
BHs collected from the literature. A fraction of these BHs are observed  
in  binaries with a nondegenerate companion, the BH masses can be dynamically measured according to the mass function
\begin{eqnarray}
f(M) \equiv \frac{P_{\rm orb}K^{3}}{2\pi G} = \frac{M_{\rm BH} \sin^{3}i}{(1+q)^2},
\end{eqnarray}
where $ P_{\rm orb} $ is the orbital period, $ K $ is the velocity semi-amplitude of the companion, $M_{\rm BH}$ is the BH mass, $ i $
is the inclination of the binary orbit, and $ q $ is the mass ratio of the companion to the BH. For a BH binary, spectroscopic
observations 
can provide the radial velocity curve of the companion star, which is able to precisely yield the 
orbital period and the radial velocity semi-amplitude  
\citep[see e.g.,][]{Casares2014,Thompson2019}. So determining the BH mass requires that the inclination and mass ratio are securely constrained. 

\begin{figure} 
   \centering
   \includegraphics[width=8.0cm, angle=0]{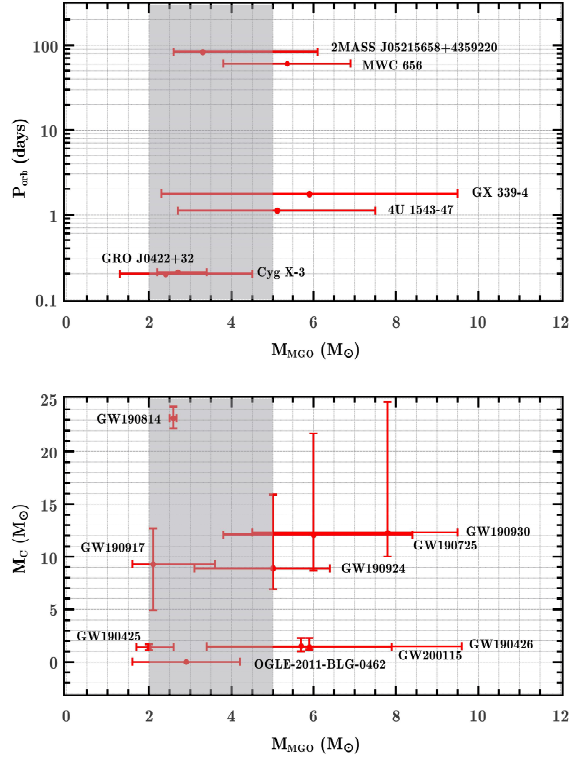}
   \caption{Top panel shows orbital period ($P_{\rm orb}$) as a function of mass ($M_{\rm MGO}$) for probable mass-gap objects
   in binaries. Bottom panel shows both component masses 
   ($M_{\rm C}$ vs. $M_{\rm MGO}$)
   for gravitational wave transients with one component falling in the mass gap. This panel also includes the compact object detected in the gravitational 
   microlensing event OGLE-2011-BLG-0462. The gray rectangle in each panel highlights the mass gap of $ 2-5M_{\odot} $.} 
   \label{Fig1}
   \end{figure}

\subsection{Low-mass X-ray binaries}

\subsubsection{4U 1543-47} 
Spectroscopic observations of 4U 1543-47 reveal a radial velocity curve with an orbital period of $ \sim 1.1  $ days
and a semi-amplitude of $ \sim 124\,\rm km\, s^{-1} $ \citep{Orosz1998}. The companion is classified as an A2 main-sequence star, 
corresponding to a mass of $ \sim 2.3-2.6M_{\odot} $. For this binary, the orbital inclination is limited to the range of $ \sim 20^{\circ}-40^{\circ} $.
The mass of the compact object is then calculated to be $ 2.7- 7.5M_{\odot}$, suggesting that it is most likely a BH \citep{Orosz1998}
and probably in the mass gap.

\subsubsection{GX 339-4} 
GX 339-4 has been studied frequently since its discovery \citep[][and references therein]{Markert1973,Heida2017}, which 
is suspected to contain a stripped giant and an accreting BH in a $ \sim 1.7 $ day orbit \citep{MD2008}. Dynamical mass of the BH reported 
by \citet{MD2008} was massive than about $ 6M_{\odot} $  \citep[see also][]{Hynes2003}. Recently, \citet{Heida2017} carried out the observations
of this source in quiescence and measured the radial velocity curve with the absorption lines from the companion star. Further data analyses revealed
a mass function of $ \sim 1.9M_{\odot} $, which lower than previously reported by a factor of about 3. The BH mass was finally constrained to 
$ 2.3-9.5M_{\odot} $, likely falling in the mass gap \citep{Heida2017} . 

\subsubsection{GRO J0422+32} 
GRO J0422+32 hosts an M-type main-sequence star and an accreting BH orbiting each other every $ \sim 0.21 $ days. The 
orbital inclination of this binary has been frequently measured, covering a wide range of $ \sim 10^{\circ}-50^{\circ} $ 
\citep[][and references therein]{Casares2014}. Note that all these inclinations are suspected since very large flickering amplitude is present 
in the optical/NIR light curves. A possible large inclination results in a lower limit of $ \sim 2.2M_{\odot} $ for the BH mass \citep{Webb2000}. 
However, the big uncertainty from the inclination means that the BH mass can reach as high as $ \gtrsim 10M_{\odot} $ \citep{Reynolds2007}. 
More recently, \citet{Casares2022} used a correlation between $ H_{\alpha} $ trough depth and inclination to derive the 
inclination of $55.6\pm 4.1^{\circ}$. Then the mass of the BH in this binary was constrained to be $2.7_{-0.5}^{+0.7}M_{\odot}$, placing it within the mass gap.

\subsection{High-mass X-ray binaries}

\subsubsection{Cyg X-3} 
Cyg X-3 is the only known X-ray binary with a Wolf-Rayet donor in our Galaxy and this system has a short orbital period of $ \sim 0.2 $ days
\citep{vanKerkwijk1992}. The mass of the compact object was estimated by \citet{Zdziarski2013} to be $ 2.4_{-1.1}^{+2.1} M_{\odot}$,
allowing for the presence of either an NS or a BH. The radio, infrared and X-ray properties of this source suggested that the compact object is
more likely a BH \citep{Zdziarski2013}. Recent data analyses support that the compact object is probably a light BH but the estimation of its real mass is subject
to many uncertainties \citep{Koljonen,Antokhin2022}.

\subsection{Detached binaries with an optical companion}

\subsubsection{MWC 656} 

MWC 656 is a Be star probably associated with the gamma-ray source AGL J2241+4454 \citep{MA2016}. The analysis of optical photometry indicates
a modulation with a period of $ \sim 60 $ days, suggesting MWC 656 is a member of a binary system \citep{Williams2010}. Using the Be-star mass from its spectral classification and the
mass ratio from the
radial velocity curves of the Be star and the companion's disk, \citet{cnr2014} 
identified the dark companion as a BH with mass of $ 3.8-6.9 M_{\odot}$. Contrary to Be/X-ray binaries with an accreting NS, 
it is interesting that the BH in this source is X-ray quiescent \citep[see also][]{Ribo2017}. However, \citet{Rivinius2022} suggested 
that MWC 656 is more likely to host a hot subdwarf \citep[i.e., an sdO+Be star binary,][]{sl2021a} instead of a BH.

\subsubsection{2MASS J05215658+4359220}

\citet{Thompson2019} combined 
radial velocity and photometric variability data to show that the giant star 2MASS J05215658+4359220 orbits a massive unseen object every $ \sim 83 $ days. Constraints on the giant's mass and radius implied that the unseen object has a mass of $ 3.3_{-0.7}^{+2.8} M_{\odot}$, suggesting it to be a low-mass BH within the mass gap. 
Follow-up X-ray observations of this source that yield an upper limit on X-ray emission indicated its nature of a noninteracting binary with a quiescent BH \citep{Thompson2019}. However, \citet{vt2020} argued that the 
unseen object can be a close binary consisting of two main-sequence stars since this can naturally explain why no X-ray emission is detected.  

\subsection{Gravitational microlensing sources}

\subsubsection{OGLE-2011-BLG-0462}

Gravitational microlensing has been proposed to detect isolated BHs in the Milky Way. When a BH lens
passes between a source and observer, there is a transient brightening of the source due to gravitational microlensing. 
Recently, OGLE-2011-BLG-0462 has been reported to be the first  discovery of a compact object with astrometric 
microlensing \citep{Sahu2022,Lam2022}. It was showed by \citet{Sahu2022} that the lens is a BH with mass of $ 7.1\pm1.3M_{\odot} $ and has a 
transverse space velocity of $ \sim 45 \,\rm km \, s^{-1} $ at a distance of $ 1.58\pm0.18 $ kpc. 
Alternatively, \citet{Lam2022} suggested that the compact-object lens is 
either an NS or a BH with mass of $ 1.6-4.2M_{\odot} $ and it has a slow transverse motion of $ <  25 \,\rm km \, s^{-1}$ at a distance of $ 0.69-1.75 $ kpc. 
Later, \citet{Mroz2022} proposed that systematic errors are a cause of the discrepant mass measurements 
and the lens should be a BH with mass of $ 7.88\pm 0.82M_{\odot} $. 
Considering the compact object to be a BH,
population synthesis simulations indicate that this BH more likely has a binary origin \citep{Andrews2022,vg2022}.

\subsection{Gravitational wave transients}

\subsubsection{GW190814 and others}

GW190814 contains a compact object of mass $ 2.59_{-0.09}^{+0.08}M_{\odot} $ that definitely lies in the mass gap \citep{Abbott2020b}. 
This compact object might be a very heavy NS 
or a very light BH \citep[e.g.,][]{Zhang2020,Huang2020,Tsokaros2020,Most2020,Godzieba2021,Bombaci2021,Zhou2021,Biswas2021}.
Besides, seven other
sources (GW190425, GW190426, GW190725, GW190917, GW190924, GW190930, GW200115) 
with false alarm rate of $ < 1\rm\, yr^{-1} $ may also host a
BH component within the mass gap \citep{lvk2021}. Interestingly, \citet{Bianconi2022} proposed that gravitational lensing is able to
delay the arrival of gravitational-wave
signals and alter their amplitudes, providing a possible explanation of the detection of mass-gap objects as gravitationally lensed mergers of binary NSs.

\section{Formation of black holes in binaries}

The general picture for the formation and evolution of isolated BH binaries has been well established over past decades \citep{vdh2009,tv2023}, including the evolutionary sequences from X-ray binaries
\citep[e.g.,][]{vdh1975,Podsiadlowski2003,Belczynski2012,Zuo2014,Fragos2015,Qin2019,sl2020} to double compact objects and gravitational wave sources
\citep[e.g.,][]{Tutukov1993,Lipunov1997,Nelemans2001,Voss2003,Kalogera2007,Belczynski2016,Eldridge2016,vdh2017,Giacobbo2018,Kruckow2018,Mapelli2018,Ablimit2018,
Breivik2020,sl2021b,Broekgaarden2021,Broekgaarden2022,vs2022}. 
In the canonical channel of forming BH systems, they are thought to stem from the primordial binaries with two zero-age main-sequence stars. 
It is required that the primordial binaries contain at least one massive
star of $ \gtrsim 15-25M_{\odot} $ for creating a BH \citep{Woosley1995,Fryer2012,Kochanek2014,Sukhbold2016,Raithel2018,Ertl2020,Mandel2020}.

Evolution of massive primordial binaries still remains some major uncertainties \citep{Podsiadlowski1992,Langer2012}, such as the process of mass transfer between binary components. 
Depending on the structure of the primary donor, the mass ratio of the binary components and how
conservative the mass transfer is, the binary systems are expected to undergo either a stable mass-transfer phase or a
dynamically unstable mass-transfer phase \citep[e.g.,][]{Soberman1997,Chen2008,Ge2010,Ge2020,sl2014,Pavlovskii2015,Pavlovskii2017,Marchant2021}. 
Compared to stable mass transfer via Roche lobe
overflow, dynamically
unstable mass transfer can trigger common-envelope evolution, during which the binary orbit dramatically shrinks 
inside a shared envelope stemming from the donor and the orbital energy of the binary system is dissipated to unbind
the common envelope \citep{Paczynski1976,Webbink1984,Iben1993,Ivanova2013}. It
is believed that the BH binaries with a 
low-mass ($\lesssim 1-2M_{\odot}$) secondary must have undergone a common-envelope phase while the ones with a
high-mass ($ \gtrsim 8-10M_{\odot} $) secondary more likely have experienced a stable mass-transfer phase \citep[e.g.,][]{Podsiadlowski2003,sl2020}. As the
evolutionary products of these BH binaries are totally different, 
I separately discuss the two cases in the following.

\subsection{The case of BH binaries with a low-mass secondary}

Figure~2 shows a schematic diagram for the formation and evolutionary track of a BH binary
with a low-mass secondary. Since the primordial binary has an extreme mass ratio, the mass
transfer between binary components is dynamically unstable and common-envelope 
evolution is unavoidable \citep[e.g.,][]{Podsiadlowski2003}. Considering that this binary can survive the
common-envelope phase, the whole hydrogen-rich envelope of the primary star is 
successfully expelled. At the same time, the secondary star is expected to accrete hardly
during the common-envelope evolution. The immediate binary hosts a massive Wolf-Rayet star
and a low-mass main-sequence star. After the Wolf-Rayet star eventually collapses into 
a BH via a supernova explosion, this binary evolves to be a BH system with a low-mass
secondary if it is not disrupted due to a supernova kick\footnote{Observations of X-ray binaries indicate that 
BHs therein possess natal kick velocities of $ \lesssim 80\,\rm km\, s^{-1} $ \citep{Mandel2016}, 
with some possible exceptions \citep[e.g.,][]{Fragos2009,Repetto2017}.}.
As a result, the initial orbit 
of this BH binary is usually eccentric. On the other hand, tidal interaction is able to circularize the
binary orbit before the low-mass companion fills its Roche lobe. The source with a low-mass giant 2MASS J05215658+4359220 
\citep{Thompson2019} is probably such a detached BH binary
formed in the canonical scenario as described above \citep[see also][]{Breivik2019,sl2020}. When Roche
lobe overflow starts from the low-mass secondary to the BH, the system enters the stage of a low-mass X-ray binary. 
During this stage, stable mass accretion can result in efficient spin-up of the BH \citep{King1999} and produce
a significant BH spin consistent with observations \citep{Podsiadlowski2003,Fragos2015,sl2020}.
It is possible that the binary always appears as a low-mass X-ray binary within a Hubble time
or evolves to be a detached BH system orbiting by a WD.

\begin{figure} 
   \centering
   \includegraphics[width=5.0cm, angle=0]{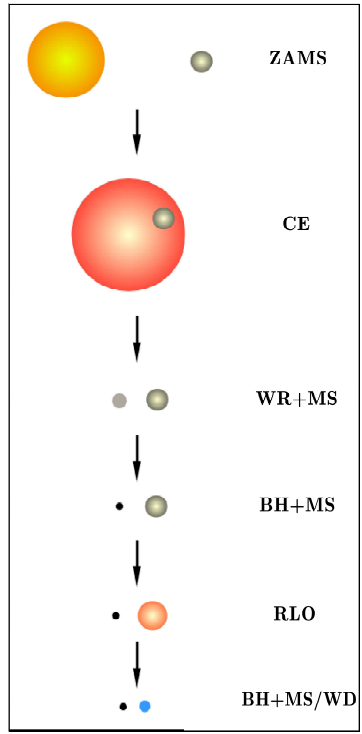}
   \caption{Illustration of the formation and evolution of a BH binary with a low-mass ($\lesssim 1-2M_{\odot}$) secondary, 
   involving the evolutionary consequences of a detached 
   system with a main-sequence companion, a low-mass X-ray binary and possibly a (detached) system with a white-dwarf companion. Acronyms used in this figure--ZAMS: zero-age main
   sequence; CE: common envelope; WR: Wolf Rayet; RLO: Roche lobe overflow; BH: black hole; WD: white dwarf.} 
   \label{Fig2}
   \end{figure}

There is a long-standing problem for the formation of the BH binaries with a low-mass
secondary \citep[see][for a review]{Li2015}. Evolved from a primordial binary, its orbital
energy may be insufficient to eject the whole envelope of the primary star during 
common-envelope evolution, given that the primary is massive than 
$ \sim 20-25M_{\odot}$ for creating a BH \citep{pz1997,Kalogera1999,Podsiadlowski2003,sl2020}.
Some exotic formation scenarios have been proposed to solve this problem, including triple star 
evolution \citep{Eggleton1986,Naoz2016}, a low-mass companion originating from a 
disrupted envelope
of the primary star \citep{Podsiadlowski1995}, the donor being a pre-main-sequence star
\citep{Ivanova2006}, the ejection of a massive envelope with additional nuclear energy
\citep{Podsiadlowski2010}, or a dynamical origin involving fly-by stars in the Galactic
field \citep{Michaely2016,Klencki2017}. Besides, it is suggested that the BH binaries with a
low-mass companion descend from those systems with an initially intermediate-mass companion as they 
are more likely to survive common-envelope evolution in the canonical formation scenario
\citep{Justham2006,Chen2006,Li2008}. In addition, a possible solution to this problem is 
enhancing the ejection efficiency for common-envelope evolution
\citep{Kiel2006,Yungelson2008}. More recent investigations showed that the BH binaries 
with a low-mass secondary can effectively produced in the canonical scenario 
if assuming the BH's progenitors (i.e., the primary stars) have relatively 
small masses of $\gtrsim 15M_{\odot}$ \citep{sl2019,sl2020}. This assumption is
consistent with some recent results according to numerical simulations of supernova explosions
\citep[e.g.,][]{Sukhbold2016,Raithel2018,Ertl2020}. Furthermore, the mass distribution of BHs in low-mass X-ray
binaries shows a rapid decline at the high-mass end of $\sim 10M_{\odot}$ \citep{ozel2010}. The narrow distribution of 
BH masses may
provide clues to constrain the evolutionary path relevant to common-envelope phases \citep{Wang2016}, although significant mass
increase of BHs is possible during the stage of X-ray binaries
\citep[e.g.,][]{Fragos2015}.

Until now, most of confirmed BH systems in the Milky Way are low-mass X-ray binaries. 
These binaries
always appear as transient sources with episodic outbursts, which are likely to be
triggered by thermal and viscous instabilities in an accretion disk around the BH 
\citep{vp1996,Dubus1999,Lasota2001,Coriat2012}. For the long-term evolution of 
low-mass X-ray binaries, there exists
a critical bifurcation orbital period of a few days \citep{Pylyser1988,Pylyser1989} to 
separate them into converging and diverging systems. It is expected that diverging systems
evolve to be wide binaries with a WD companion, while converging systems always
contract to become tight X-ray binaries with a main-sequence donor. It is possible that 
the systems with orbital periods close to the critical value evolve to be tight
BH binaries with a WD companion. These binaries may appear as detectable 
gravitational wave sources like tight NS+WD binaries that have been widely studied
\citep{Tauris2018,Chen2020,Wang2021,Deng2021,Chen2021}. Note that the real process of the
evolution of BH low-mass X-ray binaries may be very complicated. Observations indicate
that some such binaries are undergoing extremely rapid orbital decays, which are
significantly larger than expected under the conventional mechanisms of magnetic
braking and gravitational wave radiation \citep{gh2012,gh2014,gh2017}. Extra mechanisms may also work
during the evolution of BH low-mass X-ray binaries, e.g., the interaction of a binary
with its surrounding circumbinary disk \citep{Chen2015,Xu2018,Chen2019}.

\subsection{The case of BH binaries with a high-mass secondary}

Figure~3 shows a schematic diagram for the formation and evolutionary track of a BH binary
with a high-mass secondary. In this case, both components of the primordial binary are massive
stars that can eventually form BHs or NSs. During the evolution, the primary star first evolves
to fill its Roche lobe and transfers its envelope to the secondary star. The mass transfer in 
such a binary can take place stably via the process of Roche lobe overflow. After mass transfer,
the primary star loses its hydrogen envelope to become a Wolf-Rayet star and the secondary star
increases its mass due to accretion. When the Wolf-Rayet star collapses into a BH,
the immediate system is a BH binary with a high-mass
secondary (e.g., MWC 656). In this binary,
mass transfer starts as the secondary star fills its Roche lobe. As a result, the hydrogen
envelope of the secondary is likely to be stripped by the BH via either stable Roche-lobe
overflow or common-envelope evolution. The subsequent system contains a BH and another
Wolf-Rayet star (e.g., Cyg X-3). As the Wolf-Rayet star ends its life as a BH or an NS, the final binary
possesses two compact objects. If the system is tight enough, it may be observable as a
gravitational wave source. Observations of BH+BH mergers indicate some BHs with high spins
\citep{lvk2021}, who are spun up possibly via tidal interaction \citep[e.g.,][]{Qin2018,Belczynski2020,Bavera2021,Fuller2022}
 or stable mass transfer \citep{vs2020,sl2022} during progenitor evolution.

\begin{figure} 
   \centering
   \includegraphics[width=5.0cm, angle=0]{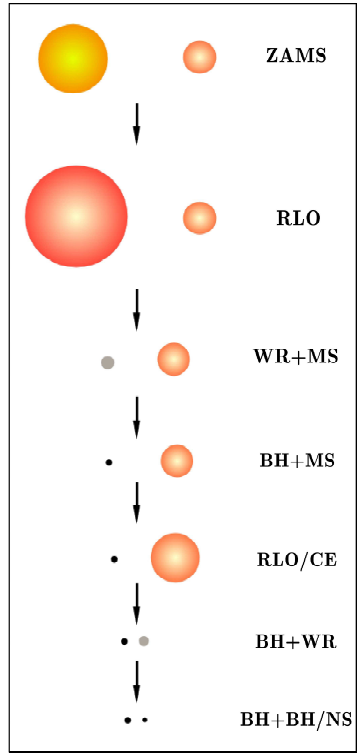}
   \caption{Illustration of the formation and evolution of a BH binary with a high-mass ($ \gtrsim 8-10M_{\odot} $) secondary, involving the evolutionary
   consequences of a detached
   system with a main-sequence star, a high-mass X-ray binary with an OB star or a Wolf-Rayet star, and finally a BH+BH or BH+NS system.} 
   \label{Fig3}
   \end{figure}

During the evolution of massive primordial binaries, 
mass accretion can rejuvenate the secondary 
star \citep{Hurley2002} and cause it to expand and spin up \citep{Neo1977,Packet1981}. In the case
of rapid mass accretion, the secondary star will get out of thermal equilibrium with significant 
expansion. This expanding secondary may even fill its own Roche lobe, leading to the formation of a 
contact system \citep{Nelson2001} and then probably a binary merger. Hence, the mass-transfer 
efficiency, i.e., the fraction of matter accreted onto the secondary among all transferred matter, is
a key factor determining whether a binary system evolves into contact. This efficiency has been 
widely discussed in the literature
\citep[e.g.,][]{Vanbeveren1979,dl1992,Wellstein2001,Petrovic2005,dm2007,sl2014,sl2016}. It is suggested
that the secondary star with rapid rotation will suppress its mass increment and drop the
mass-transfer efficiency \citep{Petrovic2005,Stancliffe2009}. It is known that the rotation evolution
of the secondary star is mainly controlled by the torques from mass accretion and tidal interaction 
\citep{dm2013,sl2014}. In wide systems tidal effects usually can be neglected, while in close systems
they tend to prevent the spin-up of the secondary star due to mass and angular momentum accretion. As
a consequence, the mass-transfer efficiency is expected to be higher than $\sim 20\%$ for 
close (Case A) binaries while drop to a level of $\sim 10\%$ or lower for wide (Case B) 
binaries \citep{sl2016,Langer2020}. This rotation-dependent picture for mass-transfer efficiencies
can well describe the formation of some massive binaries such as Wolf-Rayet star+O star systems
\citep{Petrovic2005,sl2016} and BH+Be star systems \citep{sl2014}. In this situation, the maximal mass ratio of the primary to the secondary for avoiding common-envelope
evolution can reach as high as $\sim 6$ (see Figure~4 for more details).
However, the mass 
distribution of Be stars with an NS or a subdwarf companion suggests a more higher
efficiency of $\gtrsim 0.3$ even for wide systems \citep{sl2014,Schootemeijer2018,Vinciguerra2020}. 
 
\begin{figure} 
   \centering
   \includegraphics[width=8.0cm, angle=0]{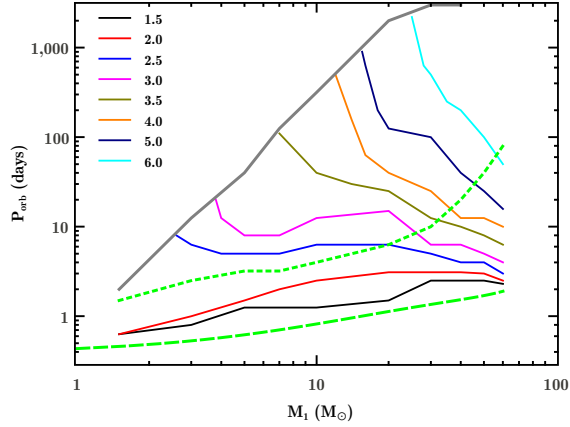}
   \caption{Allowed parameter spaces (solid curves) in the orbital period ($ P_{\rm orb} $) vs. primary mass
   ($ M_{1} $) plane for stable mass transfer in which contact and common-envelope phases are avoidable, by assuming that the mass-accretion rate onto the secondary to be 
   dependent on its rotating velocity (i.e. the mass-transfer rate multiplied by a factor of $(\Omega_{\rm K}-\Omega)/\Omega_{\rm K}$, where $\Omega$ and $\Omega_{\rm K}$ are the angular velocity of the secondary and the corresponding Keplerian limit). The 
   numbers next to each coloured label represent the mass ratio of the primary to the secondary. The green dashed, green dotted, and gray solid curves correspond to the orbital periods when the primary overflows its Roche lobe at zero-age main sequence, the end of main sequence, and the end of Hertzsprung gap, respectively. Figure from \citet{sl2014}.} 
   \label{Fig4}
   \end{figure}

\begin{figure} 
   \centering
   \includegraphics[width=15.5cm, angle=0]{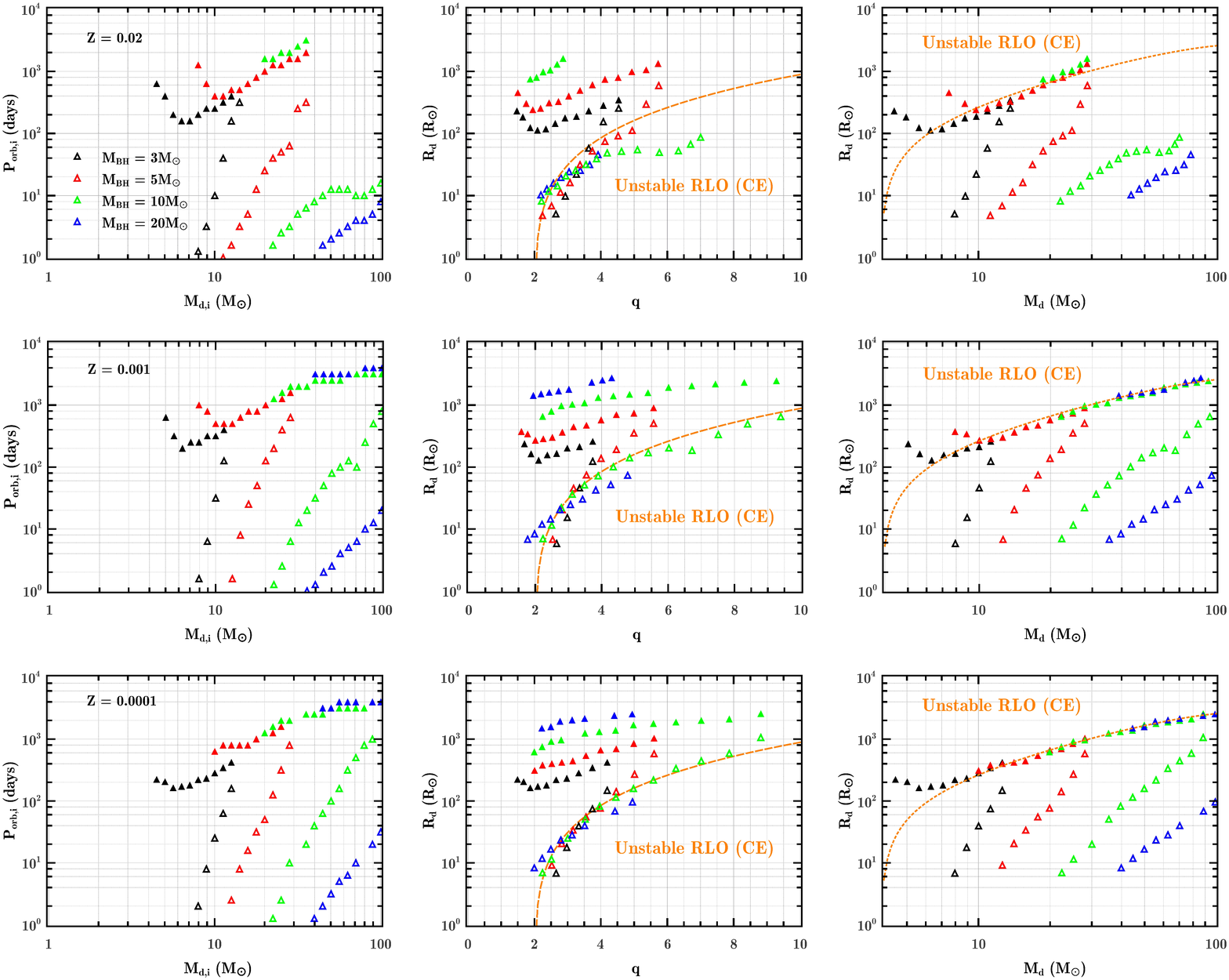}
   \caption{Parameter spaces for stable and unstable mass transfer in the BH binaries with nondegenerate donors at different metallicities ($ Z $). The left panels show the parameter space outlines distributed in the donor mass$ - $orbital period 
   ($ M_{\rm d,i}-P_{\rm orb,i} $) plane, while the middle and right panels correspond to the cases in the mass ratio$ - $donor
   radius ($ q-R_{\rm d} $) and donor mass$ - $donor radius ($ M_{\rm d} -R_{\rm d} $) planes, respectively. 
   The coloured triangles in each panel denote the calculated boundaries for the binaries with different initial BH masses 
   ($ M_{\rm BH} $), and the filled and open ones represent the upper and lower boundaries, respectively. The orange dashed curves roughly fit the lower boundaries of the donor radius as a function of the mass ratio, while the orange dotted curves for the upper boundaries as a function of the donor mass. Figure from \citet{sl2021b}.} 
   \label{Fig5}
   \end{figure}

Mass transfer stability in the binaries with an accreting BH and a massive donor has been frequently discussed
over past years, since it was applied to understand the formation channel of binary BH mergers 
\citep[e.g.,][]{vdh2017,Pavlovskii2017,Inayoshi2017,Neijssel2019,Olejak2021,Marchant2021,sl2021b,gg2021,vs2022,Dorozsmai2022}. In some cases, 
mass-transferring BH binaries with a massive donor are likely to undergo the expansion or convection
instability that followed by common-envelope evolution \citep{Pavlovskii2017}. For relatively close systems,
the expansion instability may take place when the donor stars are experiencing a period of fast
thermal-timescale
expansion. For relatively wide systems, the convection instability may happen if the donor stars have
developed a sufficiently deep convective envelope. \citet{Pavlovskii2017} suggested that there exist the
smallest radius $R_{\rm U}$ and the maximum radius $R_{\rm S}$ for which the convection and expansion
instabilities can occur, respectively. 
With detailed binary evolution simulations, \citet{sl2021b}
ran a large grid of the initial parameters for the BH binaries with a massive donor to deal with 
mass transfer stability, and obtained thorough criteria for the occurrence of common-envelope evolution. These
criteria are summarized as follows \citep[see also][]{sl2021b}. Mass transfer is always stable if the mass ratio $q$ 
of the donor to the BH is less than the minimal value $q_{\rm min}$, i.e.,
\begin{equation}
q < q_{\rm min} \sim 1.5-2.0
\end{equation}
 and always unstable if the mass ratio $q$ is
larger than the maximal value $q_{\rm max}$, i.e.,
 \begin{equation}
q > q_{\rm max} \sim 2.1+0.8M_{\rm BH}.
\end{equation}
For the systems with 
$q_{\rm min}< q < q_{\rm max}$, dynamically unstable mass transfer ensues if 
the donor radius $R_{\rm d}$ is either smaller than $R_{\rm S}$, i.e.,
\begin{equation}
R_{\rm d} < R_{\rm S} \sim 6.6-26.1q+11.4q^2
\end{equation}
or larger than $R_{\rm U}$, i.e.,
\begin{equation}
R_{\rm d} > R_{\rm U} \sim -173.8+45.5M_{\rm d}-0.18M_{\rm d}^2.
\end{equation}
Here all radii and masses are expressed in solar units. Figure~5 shows
the detailed parameter-space distributions for mass-transfer stability in the BH binaries with a nondegenerate donor. 
It can be seen that the maximal mass ratios
for avoiding common-envelope evolution are related to BH masses and there is a tendency that the 
binaries with heavier (lighter) BHs are more likely to undergo stable mass transfer (common-envelope evolution).

\subsection{Formation of BHs from accreting/merging NSs in binaries}

Mass increase of NSs may lead to their collapse into BHs, which process is likely to
take place in various types of NS binaries with a nondegenerate-star or a compact-object 
companion \citep[e.g.,][]{Vietri1999,Dermer2006,Yang2020,Gao2022,Qu2022,Siegel2022}. 
The standard picture for the formation and evolution of NS binaries is similar to the case of the BH
binaries described above. It was proposed by \citet{Gao2022} that BHs can form from
accretion-induced collapse of NSs in X-ray binaries. 
In this case, super-Eddington accretion is critical for the mass growth of NSs. Under 
this assumption, accretion-induced collapse of NSs may account for the formation
of some mass-gap BHs observed in X-ray binaries like GRO J0422+32 \citep{Gao2022}. 
This case is supported by the observations that some NSs in redback and black-widow binaries can increase their
masses to $\sim 2.4M_{\odot}$ during previous mass accretion  \citep{Linares2018,Kandel2020,Romani2022}

For the NS binaries with a high-mass
donor, dynamically unstable mass transfer seems to be unavoidable during the evolution since the
extreme mass ratios of the binary components. During a common-envelope phase, a spiral-in NS may 
suffer from hypercritical accretion and then collapse into a BH \citep[e.g.,][]{Chevalier1993,vs2020}. 
It is complicated for the process of NS accretion if it develops an accretion disk and launches jets during
the spiral-in stage \citep{Soker2018,Soker2019}.
On the other hand, recent
hydrodynamical simulations reveal that an NS embedded in a common envelope only accretes very limited 
material with mass of $\lesssim 0.1 M_{\odot}$ \citep{MacLeod2015}, which are consistent
with the mass distribution of the recycled pulsars in the tight binaries with a WD or 
an NS companion \citep[e.g.,][]{Tauris2017}. It is also possible that 
low-mass BHs are the remnants of binary NS mergers like GW170817 \citep{Abbott2017}
and GW190425 \citep{Abbott2020a}. 
There is a common feature that the cases of NS accretion and binary mergers are more likely to produce
low-mass BHs within the mass gap.

\section{Prospects of searching for black holes}

In this section, I mainly focus on a series of our recent works that assessing the detectability of 
various types of BH binaries including the detached systems with a main-sequence/giant 
companion \citep{sl2019}, high-/low-mass X-ray binaries \citep{sl2020}, the systems with a pulsar
companion \citep{sl2018a}, and the tight systems with a WD/NS/BH companion as gravitational 
wave sources \citep{sl2021b}. Table 1 shows the estimated birthrates and numbers of these types of BH binaries in the Milky Way. 

\begin{table}
\bc
\begin{minipage}[]{150mm}
\caption[]{Estimated birthrates and numbers for various types of BH binaries in the Milky Way, by assuming an averaged star formation rate of $\sim 3 M_{\odot}\rm yr^{-1}$ over the past 10 Gyr. \label{tab1}}\end{minipage}
\setlength{\tabcolsep}{15.0pt}
\small
 \begin{tabular}{lccccccccccc}
  \hline\noalign{\smallskip}
Binary types & $R_{\rm birth}$ ($\rm Myr^{-1}$) &  $N_{\rm total}$ & $N_{\rm detect}$   \\
  \hline\noalign{\smallskip}
BH+MS & $45-130$ & $470-12000$ & $260-930^{a}$  \\
BH+G & $-$ & $2-600$ & $ 2-50 ^{a}$ \\
RLO XRB & $-$ & $50-820$ & $-$ \\
BH+OB & $-$ & $110-440$ & $10-30^{b}$ \\
BH+Be & $-$ & $330-1000$ & $-$ \\
BH+He & $-$ & $200-540$ & $30-110^{b}$ \\
BH+PSR & $1-10$ & $3-80$ & $<8^{c}$\\
BH+BH & $20-150$ &  $10^5-10^6$ & $12-26^{d}$ \\ 
BH+NS & $4-30$ & $10^4-10^5$ & $2-14^{d}$\\
BH+WD & $10-100$ & $10^5-10^6$ & $<38^{d}$ \\
  \noalign{\smallskip}\hline
\end{tabular}
\ec
\tablecomments{0.99\textwidth}{$N_{\rm total}$ and $N_{\rm detect}$ denote the total and detectable 
numbers of different types of Galactic BH binaries, respectively. The uncertainties of
these numbers are mainly originated from the assumptions of different supernova models. 
$^{a}$BH+MS/BH+G represent detached
BH systems with a main-sequence/giant companion, assuming they can be detected by Gaia if the optical companion 
is brighter than 20 mag \citep{sl2019}. $^{b}$BH+OB/BH+He represent wind-fed X-ray binaries
with a regular OB-star/helium-star companion, assuming they can be detected if their X-ray luminosities are
larger $10^{35}\rm\,erg\,s^{-1}$ \citep{sl2020}. 
$^{c}$BH+PSR represents the BH systems with a pulsar companion,
assuming they can be detected by FAST with the flux density limit of 0.005 mJy in the telescope's 
visible sky \citep{sl2018a}. 
$^{d}$BH+BH/BH+NS/BH+WD represent the BH systems with a BH/NS/WD companion, 
assuming they can be detected by LISA if the signal-to-noise ratio is larger than 5 \citep{sl2021b}.}
\end{table}

\subsection{Detached BH systems with a main-sequence/giant companion}

Optical observations of visible stars in binaries with unseen objects have been proposed to 
hunt BHs \citep{Guseinov1966,Trimble1969,Gu2019}. Identification of BHs by measuring their dynamical masses
in this way has become a hot topic \citep[e.g.,][]{Gould2002,Breivik2017,Mashian2017,Yamaguchi2018,Yalinewich2018,Andrews2019,sl2019,Yi2019,
Wiktorowicz2020,Shikauchi2020,Shikauchi2022,Janssens2022,Chawla2022}. In part of detached binaries, BHs are accreting at a sufficiently low
rate and therefore emitting undetectable X-rays. To date, a handful of binaries have been claimed to
harbour a quiescent BH \citep[e.g.,][]{Liu2019,Thompson2019,Jayasinghe2021,Saracino2022,Shenar2022b,ebr2022}. However,
it is argued that most of these sources have been shown to be incorrect classifications and  
do not contain a BH \citep[e.g.,][]{Shenar2020,vt2020,eb2022,ebs2022}. 

Although the searches of non-interacting BHs are still challenging, there must have a number of 
detached BH binaries in the Milky Way. Using the method of binary population synthesis, \citet{sl2019}
simulated the Galactic population of detached BH binaries with a main-sequence/giant companion. 
According to the conventional wisdom, BHs evolve form stars of massive than $\sim 20-25M_{\odot}$. As
I have mentioned before, it is difficult to produce the BH systems with a low-mass companion in this case,
since previous common-envelope phases always led to binary mergers. Adopting some recent suggestions 
\citep[e.g.,][]{Sukhbold2016,Raithel2018} that the masses of BH's progenitors can reach as low as 
$\sim 15M_{\odot}$ and the remnant BHs have the masses of 
pre-supernova stars, \citet{sl2019} estimated that $\sim 4000-12000$ detached BH systems can be produced in 
the Milky Way and hundreds of them are potentially observable systems by Gaia. Among all these detached 
BH binaries, the majority of them are expected to have a main-sequence companion 
while a small part of them contain a giant companion. 
Most notably, the BHs formed in this scenario tend to have
relatively large masses of $\gtrsim 5M_{\odot}$ \citep{sl2019}, allowing a presence of the mass gap.

\subsection{BH X-ray binaries}

Since the discovery of the first X-ray source Scorpius X-1 outside the solar system \citep{Giacconi1962},
there are more than 300 X-ray binaries known in the Milky Way \citep{Liu2006,Liu2007}. For a significant
fraction of X-ray binaries, the nature of accreting compact objects is unclear. Further data 
analyses and deep observations may reveal the properties of some hidden/candidate BHs 
in these existing X-ray binaries \citep[e.g.,][]{Atri2022,DB2022}.
For known BH X-ray binaries, most of them are lobe-filling systems with a low-mass companion while only
a few are wind-fed systems with a high-mass companion. In the latter case, the high-mass companion could
be an OB (supergiant) star \citep[e.g., Cyg X-1,][]{Orosz2011}, a Be star 
\citep[e.g., MWC 656,][]{Casares2014}, 
or a Wolf-Rayet (helium) star \citep[e.g., Cyg X-3,][]{Zdziarski2013}. 
Although these wind-fed binaries are detached, they may appear as bright and detectable
X-ray sources if BHs are efficiently capturing companion's winds. 

Based on the work of \citet{sl2019} for detached BH systems, 
\citet{sl2020} further followed the evolutionary tracks of the binaries and obtained the potential
population of Galactic BH X-ray binaries. Adopting a range of input physical models for BH formation,
it is estimated that the total number of the wind-fed systems with an OB star in the Milky Way is
$\sim 500-1500$, of which $\sim 3/4$ have a Be star \citep[assuming its rotational velocity larger than $80\%$
of the Keplerian limit,][]{Porter2003} and $\sim 1/4$ have a regular OB star (assuming its rotational velocity 
less than $80\%$ of the Keplerian limit). In addition, there are totally $\sim 200 -500$ binaries with a 
BH and a helium star in the Milky Way. Among all wind-fed systems, only a small fraction are 
expected to be observable X-ray binaries. For the lobe-filling systems with a low-mass donor, it is 
estimated that their total number in the Milky Way is about several hundred. 

\subsection{BH binaries with a pulsar companion}

More than forty years have passed since the discovery of the first binary pulsar B1913+16
\citep{Hulse1975}, and there are about two dozens known in the Milky Way \citep[e.g.,][]{Tauris2017,sl2018b}.
However, no binary systems with a BH and a pulsar have been detected so far. Detection of such binaries
is a holy grail in astrophysics, not only because they can provide clues to constrain the evolution
of massive binaries, but also they are very useful for testing relativistic gravity and possible new physics 
\citep[e.g.,][]{Liu2014,Seymour2018,Seymour2020,Ding2021,LL2021,Su2021,Tong2022}. 

Using a population synthesis method, \citet{sl2018a} modelled the formation of the binaries with a BH and 
an NS in the Galactic disk and obtained the birthrate of such systems to be $\sim 1-10\,\rm Myr^{-1}$. 
Considering that NSs can be detected as radio pulsars only when they are still active and beamed toward the
Earth, it is estimated that there exist $\sim 3-80$ systems with a BH and a pulsar in the Galactic disk
and around $10\%$ of them could be observed by FAST \citep{Nan2011} in the telescope's 
visible sky \citep{sl2018a}. Although these binaries are very rare, they are more likely to host a  
mass-gap BH in some specific supernova models \citep{sl2021b}.

\subsection{Gravitational wave sources containing at least a BH component}

Mergers of BH+BH or BH+NS binaries may emit observable gravitational wave signals, which can be used 
to explore the supernova mechanisms that forms NSs/BHs and test the existence of mass gap 
\citep[e.g.,][]{sl2021b,Wagg2021,Dabrowny2021,Farah2022,Olejak2022}. Most notably, the detection of GW190814 with 
a mass-gap object challenges the standard paradigm of massive binary evolution and compact-object formation
\citep[e.g.,][]{Zevin2020,Antoniadis2022}. Consequently, this mass-gap object has been 
proposed to be the result of NS accretion/mergers in active galactic nucleus disks \citep{Yang2020}, 
globular clusters \citep{Kritos2021} or hierarchical triple systems \citep{Lu2021,LiuLai2021,Cholis2022}. 
Current gravitational wave observations still identified a relative
dearth of binaries with component masses in the mass gap \citep{lvk2021}. 

To study the impact of a possible mass gap on the population properties of merging BH binaries with a 
WD/NS/BH companion, \citet{sl2021b} adopted three different supernova mechanisms to treat the compact
remnant masses including the rapid \citep{Fryer2012}, the delayed \citep{Fryer2012} and the stochastic 
\citep{Mandel2020} recipes. The rapid supernova mechanism can naturally lead to a mass gap, while the delayed
and stochastic mechanisms allow the formation of compact objects within the mass gap. With the calculations of
binary population
synthesis, \citet{sl2021b} estimated that the local merger rate density of all binaries with 
at least a BH component is about $60-200\,\rm Gpc^{-3}yr^{-1}$. In the Milky Way,  dozens of
merging BH systems are detectable by future space-based gravitational wave detectors, e.g., 
LISA \citep{as2017}, TianQin \citep{Luo2016} and Taiji \citep{Ruan2020}. 
Compared to no low-mass BHs formed via rapid supernova 
mechanism, both delayed and stochastic mechanisms predicted that $\sim 100\%/ \sim70\%/\sim30\%$ of merging BH+WD/BH+NS/BH+BH binaries are likely to have BH components within the mass gap \citep{sl2021b}. 

\subsection{Single BHs}

Besides binaries, gravitational microlensing observations of isolated BHs provide a golden opportunity for testing the existence of mass gap and
constraining the mechanism of supernova explosions \citep[e.g.,][]{Lam2020}.
On the one hand, population synthesis simulations reveal that the vast majority of Galactic BHs appear as singles \citep{Wiktorowicz2019,Olejak2020}.
On the other hand, supernova kicks are capable to disrupt the binaries with low-mass BHs \citep{Mandel2021} 
and accretion of NSs
from a companion may result in the formation of low-mass BHs in binaries \citep[e.g.,][]{vs2020,Gao2022}.

\normalem
\begin{acknowledgements}
This work was supported by the Natural Science Foundation 
of China (grant No. 11973026) and the National Program on Key Research and Development 
Project (grant No. 2021YFA0718500).

\end{acknowledgements}
  

\begin{thebibliography}{500}


\bibitem[Abbott et al.(2016)]{Abbott2016} Abbott, B.~P., Abbott, R., Abbott, T.~D., et al.\ 2016, \prl, 116, 061102

\bibitem[Abbott et al.(2017)]{Abbott2017} Abbott, B.~P., Abbott, R., Abbott, T.~D., et al.\ 2017, \prl, 119, 161101

\bibitem[Abbott et al.(2019)]{Abbott2019} Abbott, B.~P., Abbott, R., Abbott, T.~D., et al.\ 2019, Physical Review X, 9, 031040

\bibitem[Abbott et al.(2020a)]{Abbott2020a} Abbott, B.~P., Abbott, R., Abbott, T.~D., et al.\ 2020a, \apjl, 892, L3

\bibitem[Abbott et al.(2020b)]{Abbott2020b} Abbott, R., Abbott, T.~D., Abraham, S., et al.\ 2020b, \apjl, 896, L44

\bibitem[Abbott et al.(2021)]{Abbott2021} Abbott, R., Abbott, T.~D., Abraham, S., et al.\ 2021, Physical Review X, 11, 021053

\bibitem[Ablimit \& Maeda(2018)]{Ablimit2018} Ablimit, I. \& Maeda, K.\ 2018, \apj, 866, 151

\bibitem[Alsing et al.(2018)]{Alsing2018} Alsing, J., Silva, H.~O., \& Berti, E.\ 2018, \mnras, 478, 1377

\bibitem[Amaro-Seoane et al.(2017)]{as2017} Amaro-Seoane, P., Audley, H., Babak, S., et al.\ 2017, arXiv:1702.00786

\bibitem[Andrews et al.(2019)]{Andrews2019} Andrews, J.~J., Breivik, K., \& Chatterjee, S.\ 2019, \apj, 886, 68

\bibitem[Andrews \& Kalogera(2022)]{Andrews2022} Andrews, J.~J. \& Kalogera, V.\ 2022, \apj, 930, 159

\bibitem[Antokhin et al.(2022)]{Antokhin2022} Antokhin, I.~I., Cherepashchuk, A.~M., Antokhina, E.~A., et al.\ 2022, \apj, 926, 123

\bibitem[Antoniadis et al.(2013)]{Antoniadis2013} Antoniadis, J., Freire, P.~C.~C., Wex, N., et al.\ 2013, Science, 340, 448

\bibitem[Antoniadis et al.(2022)]{Antoniadis2022} Antoniadis, J., Aguilera-Dena, D.~R., Vigna-G{\'o}mez, A., et al.\ 2022, \aap, 657, L6

\bibitem[Atri et al.(2022)]{Atri2022} Atri, P., Miller-Jones, J.~C.~A., Bahramian, A., et al.\ 2022, arXiv:2209.05437

\bibitem[Bailyn et al.(1998)]{Bailyn1998} Bailyn, C.~D., Jain, R.~K., Coppi, P., et al.\ 1998, \apj, 499, 367

\bibitem[Bavera et al.(2021)]{Bavera2021} Bavera, S.~S., Fragos, T., Zevin, M., et al.\ 2021, \aap, 647, A153

\bibitem[Belczynski et al.(2012)]{Belczynski2012} Belczynski, K., Wiktorowicz, G., Fryer, C.~L., et al.\ 2012, \apj, 757, 91

\bibitem[Belczynski et al.(2016)]{Belczynski2016} Belczynski, K., Holz, D.~E., Bulik, T., et al.\ 2016, \nat, 534, 512

\bibitem[Belczynski et al.(2020)]{Belczynski2020} Belczynski, K., Klencki, J., Fields, C.~E., et al.\ 2020, \aap, 636, A104

\bibitem[Bennett et al.(2002)]{Bennett2002} Bennett, D.~P., Becker, A.~C., Quinn, J.~L., et al.\ 2002, \apj, 579, 639

\bibitem[Bianconi et al.(2022)]{Bianconi2022} Bianconi, M., Smith, G.~P., Nicholl, M., et al.\ 2022, arXiv:2204.12978

\bibitem[Biswas et al.(2021)]{Biswas2021} Biswas, B., Nandi, R., Char, P., et al.\ 2021, \mnras, 505, 1600

\bibitem[Bombaci et al.(2021)]{Bombaci2021} Bombaci, I., Drago, A., Logoteta, D., et al.\ 2021, \prl, 126, 162702

\bibitem[Breivik et al.(2017)]{Breivik2017} Breivik, K., Chatterjee, S., \& Larson, S.~L.\ 2017, \apjl, 850, L13

\bibitem[Breivik et al.(2019)]{Breivik2019} Breivik, K., Chatterjee, S., \& Andrews, J.~J.\ 2019, \apjl, 878, L4

\bibitem[Breivik et al.(2020)]{Breivik2020} Breivik, K., Coughlin, S., Zevin, M., et al.\ 2020, \apj, 898, 71

\bibitem[Broekgaarden et al.(2021)]{Broekgaarden2021} Broekgaarden, F.~S., Berger, E., Neijssel, C.~J., et al.\ 2021, \mnras, 508, 5028

\bibitem[Broekgaarden et al.(2022)]{Broekgaarden2022} Broekgaarden, F.~S., Berger, E., Stevenson, S., et al.\ 2022, \mnras

\bibitem[Brown \& Bethe(1994)]{Brown1994} Brown, G.~E. \& Bethe, H.~A.\ 1994, \apj, 423, 659

\bibitem[Casares \& Jonker(2014)]{Casares2014} Casares, J. \& Jonker, P.~G.\ 2014, \ssr, 183, 223

\bibitem[Casares et al.(2014)]{cnr2014} Casares, J., Negueruela, I., Rib{\'o}, M., et al.\ 2014, \nat, 505, 378

\bibitem[Casares et al.(2022)]{Casares2022} Casares, J., Mu{\~n}oz-Darias, T., Torres, M.~A.~P., et al.\ 2022, \mnras, 516, 2023

\bibitem[Chawla et al.(2022)]{Chawla2022} Chawla, C., Chatterjee, S., Breivik, K., et al.\ 2022, \apj, 931, 107

\bibitem[Chen \& Han(2008)]{Chen2008} Chen, X. \& Han, Z.\ 2008, \mnras, 387, 1416

\bibitem[Chen \& Li(2006)]{Chen2006} Chen, W.-C. \& Li, X.-D.\ 2006, \mnras, 373, 305

\bibitem[Chen \& Li(2015)]{Chen2015} Chen, W.-C. \& Li, X.-D.\ 2015, \aap, 583, A108

\bibitem[Chen \& Podsiadlowski(2019)]{Chen2019} Chen, W.-C. \& Podsiadlowski, P.\ 2019, \apjl, 876, L11

\bibitem[Chen et al.(2020)]{Chen2020} Chen, W.-C., Liu, D.-D., \& Wang, B.\ 2020, \apjl, 900, L8

\bibitem[Chen et al.(2021)]{Chen2021} Chen, H.-L., Tauris, T.~M., Han, Z., et al.\ 2021, \mnras, 503, 3540

\bibitem[Chevalier(1993)]{Chevalier1993} Chevalier, R.~A.\ 1993, \apjl, 411, L33

\bibitem[Cholis et al.(2022)]{Cholis2022} Cholis, I., Kritos, K., \& Garfinkle, D.\ 2022, \prd, 105, 123022

\bibitem[Clark et al.(2002)]{Clark2002} Clark, J.~S., Goodwin, S.~P., Crowther, P.~A., et al.\ 2002, \aap, 392, 909

\bibitem[Coriat et al.(2012)]{Coriat2012} Coriat, M., Fender, R.~P., \& Dubus, G.\ 2012, \mnras, 424, 1991

\bibitem[Cromartie et al.(2020)]{Cromartie2020} Cromartie, H.~T., Fonseca, E., Ransom, S.~M., et al.\ 2020, Nature Astronomy, 4, 72

\bibitem[Dabrowny et al.(2021)]{Dabrowny2021} Dabrowny, M., Giacobbo, N., \& Gerosa, D.\ 2021, Rendiconti Lincei. Scienze Fisiche e Naturali, 32, 665

\bibitem[Dashwood Brown et al.(2022)]{DB2022} Dashwood Brown, C., Gandhi, P., \& Charles, P.\ 2022, arXiv:2209.09920

\bibitem[De Loore \& De Greve(1992)]{dl1992} De Loore, C. \& De Greve, J.~P.\ 1992, \aaps, 94, 453

\bibitem[de Mink et al.(2007)]{dm2007} de Mink, S.~E., Pols, O.~R., \& Hilditch, R.~W.\ 2007, \aap, 467, 1181

\bibitem[de Mink et al.(2013)]{dm2013} de Mink, S.~E., Langer, N., Izzard, R.~G., et al.\ 2013, \apj, 764, 166

\bibitem[Demorest et al.(2010)]{Demorest2010} Demorest, P.~B., Pennucci, T., Ransom, S.~M., et al.\ 2010, \nat, 467, 1081

\bibitem[Deng et al.(2021)]{Deng2021} Deng, Z.-L., Li, X.-D., Gao, Z.-F., et al.\ 2021, \apj, 909, 174

\bibitem[Dermer \& Atoyan(2006)]{Dermer2006} Dermer, C.~D. \& Atoyan, A.\ 2006, \apjl, 643, L13

\bibitem[Ding et al.(2021)]{Ding2021} Ding, Q., Tong, X., \& Wang, Y.\ 2021, \apj, 908, 78

\bibitem[Dorozsmai \& Toonen(2022)]{Dorozsmai2022} Dorozsmai, A. \& Toonen, S.\ 2022, arXiv:2207.08837

\bibitem[Dubus et al.(1999)]{Dubus1999} Dubus, G., Lasota, J.-P., Hameury, J.-M., et al.\ 1999, \mnras, 303, 139

\bibitem[Eggleton \& Verbunt(1986)]{Eggleton1986} Eggleton, P.~P. \& Verbunt, F.\ 1986, \mnras, 220, 13P

\bibitem[El-Badry \& Burdge(2022)]{eb2022} El-Badry, K. \& Burdge, K.~B.\ 2022, \mnras, 511, 24

\bibitem[El-Badry et al.(2022a)]{ebs2022} El-Badry, K., Seeburger, R., Jayasinghe, T., et al.\ 2022a, \mnras, 512, 5620

\bibitem[El-Badry et al.(2022b)]{ebr2022} El-Badry, K., Rix, H.-W., Quataert, E., et al.\ 2022b, arXiv:2209.06833

\bibitem[Eldridge \& Stanway(2016)]{Eldridge2016} Eldridge, J.~J. \& Stanway, E.~R.\ 2016, \mnras, 462, 3302

\bibitem[Ertl et al.(2020)]{Ertl2020} Ertl, T., Woosley, S.~E., Sukhbold, T., et al.\ 2020, \apj, 890, 51

\bibitem[Farah et al.(2022)]{Farah2022} Farah, A., Fishbach, M., Essick, R., et al.\ 2022, \apj, 931, 108

\bibitem[Farr et al.(2011)]{Farr2011} Farr, W.~M., Sravan, N., Cantrell, A., et al.\ 2011, \apj, 741, 103

\bibitem[Farr \& Chatziioannou(2020)]{Farr2020} Farr, W.~M. \& Chatziioannou, K.\ 2020, Research Notes of the American Astronomical Society, 4, 65

\bibitem[Fragos et al.(2009)]{Fragos2009} Fragos, T., Willems, B., Kalogera, V., et al.\ 2009, \apj, 697, 1057

\bibitem[Fragos \& McClintock(2015)]{Fragos2015} Fragos, T. \& McClintock, J.~E.\ 2015, \apj, 800, 17

\bibitem[Fryer et al.(2012)]{Fryer2012} Fryer, C.~L., Belczynski, K., Wiktorowicz, G., et al.\ 2012, \apj, 749, 91

\bibitem[Fryer et al.(2022)]{Fryer2022} Fryer, C.~L., Olejak, A., \& Belczynski, K.\ 2022, \apj, 931, 94

\bibitem[Fu et al.(2022)]{Fu2022} Fu, J.-B., Gu, W.-M., Zhang, Z.-X., et al.\ 2022, arXiv:2207.05434

\bibitem[Fuller \& Lu(2022)]{Fuller2022} Fuller, J. \& Lu, W.\ 2022, \mnras, 511, 3951

\bibitem[Gaia Collaboration et al.(2022)]{Gaia2022} Gaia Collaboration, Arenou, F., Babusiaux, C., et al.\ 2022, arXiv:2206.05595

\bibitem[Gallegos-Garcia et al.(2021)]{gg2021} Gallegos-Garcia, M., Berry, C.~P.~L., Marchant, P., et al.\ 2021, \apj, 922, 110

\bibitem[Gao et al.(2022)]{Gao2022} Gao, S.-J., Li, X.-D., \& Shao, Y.\ 2022, \mnras, 514, 1054

\bibitem[Ge et al.(2010)]{Ge2010} Ge, H., Hjellming, M.~S., Webbink, R.~F., et al.\ 2010, \apj, 717, 724

\bibitem[Ge et al.(2020)]{Ge2020} Ge, H., Webbink, R.~F., \& Han, Z.\ 2020, \apjs, 249, 9

\bibitem[Giacconi et al.(1962)]{Giacconi1962} Giacconi, R., Gursky, H., Paolini, F.~R., et al.\ 1962, \prl, 9, 439

\bibitem[Giacobbo \& Mapelli(2018)]{Giacobbo2018} Giacobbo, N. \& Mapelli, M.\ 2018, \mnras, 480, 2011

\bibitem[Giesers et al.(2018)]{Giesers2018} Giesers, B., Dreizler, S., Husser, T.-O., et al.\ 2018, \mnras, 475, L15

\bibitem[Giesers et al.(2019)]{Giesers2019} Giesers, B., Kamann, S., Dreizler, S., et al.\ 2019, \aap, 632, A3

\bibitem[Godzieba et al.(2021)]{Godzieba2021} Godzieba, D.~A., Radice, D., \& Bernuzzi, S.\ 2021, \apj, 908, 122

\bibitem[Gomel et al.(2022)]{Gomel2022} Gomel, R., Mazeh, T., Faigler, S., et al.\ 2022, arXiv:2206.06032

\bibitem[Gonz{\'a}lez Hern{\'a}ndez et al.(2012)]{gh2012} Gonz{\'a}lez Hern{\'a}ndez, J.~I., Rebolo, R., \& Casares, J.\ 2012, \apjl, 744, L25

\bibitem[Gonz{\'a}lez Hern{\'a}ndez et al.(2014)]{gh2014} Gonz{\'a}lez Hern{\'a}ndez, J.~I., Rebolo, R., \& Casares, J.\ 2014, \mnras, 438, L21

\bibitem[Gonz{\'a}lez Hern{\'a}ndez et al.(2017)]{gh2017} Gonz{\'a}lez Hern{\'a}ndez, J.~I., Su{\'a}rez-Andr{\'e}s, L., Rebolo, R., et al.\ 2017, \mnras, 465, L15

\bibitem[Gould \& Salim(2002)]{Gould2002} Gould, A. \& Salim, S.\ 2002, \apj, 572, 944

\bibitem[Gu et al.(2019)]{Gu2019} Gu, W.-M., Mu, H.-J., Fu, J.-B., et al.\ 2019, \apjl, 872, L20

\bibitem[Guseinov \& Zel'dovich(1966)]{Guseinov1966} Guseinov, O.~K. \& Zel'dovich, Y.~B.\ 1966, \sovast, 10, 251

\bibitem[Han et al.(2020)]{Han2020} Han, Z.-W., Ge, H.-W., Chen, X.-F., et al.\ 2020, Research in Astronomy and Astrophysics, 20, 161

\bibitem[Heida et al.(2017)]{Heida2017} Heida, M., Jonker, P.~G., Torres, M.~A.~P., et al.\ 2017, \apj, 846, 132

\bibitem[Huang et al.(2020)]{Huang2020} Huang, K., Hu, J., Zhang, Y., et al.\ 2020, \apj, 904, 39

\bibitem[Hulse \& Taylor(1975)]{Hulse1975} Hulse, R.~A. \& Taylor, J.~H.\ 1975, \apjl, 195, L51

\bibitem[Hurley et al.(2002)]{Hurley2002} Hurley, J.~R., Tout, C.~A., \& Pols, O.~R.\ 2002, \mnras, 329, 897

\bibitem[Hynes et al.(2003)]{Hynes2003} Hynes, R.~I., Steeghs, D., Casares, J., et al.\ 2003, \apjl, 583, L95

\bibitem[Iben \& Livio(1993)]{Iben1993} Iben, I. \& Livio, M.\ 1993, \pasp, 105, 1373

\bibitem[Inayoshi et al.(2017)]{Inayoshi2017} Inayoshi, K., Hirai, R., Kinugawa, T., et al.\ 2017, \mnras, 468, 5020

\bibitem[Ivanova(2006)]{Ivanova2006} Ivanova, N.\ 2006, \apjl, 653, L137

\bibitem[Ivanova et al.(2013)]{Ivanova2013} Ivanova, N., Justham, S., Chen, X., et al.\ 2013, \aapr, 21, 59

\bibitem[Janssens et al.(2022)]{Janssens2022} Janssens, S., Shenar, T., Sana, H., et al.\ 2022, \aap, 658, A129

\bibitem[Jayasinghe et al.(2021)]{Jayasinghe2021} Jayasinghe, T., Stanek, K.~Z., Thompson, T.~A., et al.\ 2021, \mnras, 504, 2577

\bibitem[Jayasinghe et al.(2022)]{Jayasinghe2022} Jayasinghe, T., Rowan, D.~M., Thompson, T.~A., et al.\ 2022, arXiv:2207.05086

\bibitem[Justham et al.(2006)]{Justham2006} Justham, S., Rappaport, S., \& Podsiadlowski, P.\ 2006, \mnras, 366, 1415

\bibitem[Kaczmarek et al.(2022)]{Kaczmarek2022} Kaczmarek, Z., McGill, P., Evans, N.~W., et al.\ 2022, \mnras, 514, 4845

\bibitem[Kalogera(1999)]{Kalogera1999} Kalogera, V.\ 1999, \apj, 521, 723

\bibitem[Kalogera et al.(2007)]{Kalogera2007} Kalogera, V., Belczynski, K., Kim, C., et al.\ 2007, \physrep, 442, 75

\bibitem[Kandel \& Romani(2020)]{Kandel2020} Kandel, D. \& Romani, R.~W.\ 2020, \apj, 892, 101

\bibitem[Kiel \& Hurley(2006)]{Kiel2006} Kiel, P.~D. \& Hurley, J.~R.\ 2006, \mnras, 369, 1152

\bibitem[King \& Kolb(1999)]{King1999} King, A.~R. \& Kolb, U.\ 1999, \mnras, 305, 654

\bibitem[Klencki et al.(2017)]{Klencki2017} Klencki, J., Wiktorowicz, G., G{\l}adysz, W., et al.\ 2017, \mnras, 469, 3088

\bibitem[Kobulnicky et al.(2014)]{Kobulnicky2014} Kobulnicky, H.~A., Kiminki, D.~C., Lundquist, M.~J., et al.\ 2014, \apjs, 213, 34

\bibitem[Kochanek(2014)]{Kochanek2014} Kochanek, C.~S.\ 2014, \apj, 785, 28

\bibitem[Koljonen \& Maccarone(2017)]{Koljonen} Koljonen, K.~I.~I. \& Maccarone, T.~J.\ 2017, \mnras, 472, 2181

\bibitem[Kreidberg et al.(2012)]{Kreidberg2012} Kreidberg, L., Bailyn, C.~D., Farr, W.~M., et al.\ 2012, \apj, 757, 36

\bibitem[Kritos \& Cholis(2021)]{Kritos2021} Kritos, K. \& Cholis, I.\ 2021, \prd, 104, 043004

\bibitem[Kroupa et al.(1993)]{Kroupa1993} Kroupa, P., Tout, C.~A., \& Gilmore, G.\ 1993, \mnras, 262, 545

\bibitem[Kruckow et al.(2018)]{Kruckow2018} Kruckow, M.~U., Tauris, T.~M., Langer, N., et al.\ 2018, \mnras, 481, 1908

\bibitem[Lam et al.(2020)]{Lam2020} Lam, C.~Y., Lu, J.~R., Hosek, M.~W., et al.\ 2020, \apj, 889, 31

\bibitem[Lam et al.(2022)]{Lam2022} Lam, C.~Y., Lu, J.~R., Udalski, A., et al.\ 2022, \apjl, 933, L23

\bibitem[Langer(2012)]{Langer2012} Langer, N.\ 2012, \araa, 50, 107

\bibitem[Langer et al.(2020)]{Langer2020} Langer, N., Sch{\"u}rmann, C., Stoll, K., et al.\ 2020, \aap, 638, A39

\bibitem[Lasota(2001)]{Lasota2001} Lasota, J.-P.\ 2001, \nar, 45, 449

\bibitem[Li(2008)]{Li2008} Li, X.-D.\ 2008, \mnras, 384, L16

\bibitem[Li(2015)]{Li2015} Li, X.-D.\ 2015, \nar, 64, 1

\bibitem[Li et al.(2021)]{Li2021} Li, Y.-J., Tang, S.-P., Wang, Y.-Z., et al.\ 2021, \apj, 923, 97

\bibitem[Linares et al.(2018)]{Linares2018} Linares, M., Shahbaz, T., \& Casares, J.\ 2018, \apj, 859, 54

\bibitem[Lipunov et al.(1997)]{Lipunov1997} Lipunov, V.~M., Postnov, K.~A., \& Prokhorov, M.~E.\ 1997, \mnras, 288, 245

\bibitem[Liu \& Lai(2021)]{LiuLai2021} Liu, B. \& Lai, D.\ 2021, \mnras, 502, 2049

\bibitem[Liu et al.(2019)]{Liu2019} Liu, J., Zhang, H., Howard, A.~W., et al.\ 2019, \nat, 575, 618

\bibitem[Liu et al.(2014)]{Liu2014} Liu, K., Eatough, R.~P., Wex, N., et al.\ 2014, \mnras, 445, 3115

\bibitem[Liu et al.(2006)]{Liu2006} Liu, Q.~Z., van Paradijs, J., \& van den Heuvel, E.~P.~J.\ 2006, \aap, 455, 1165

\bibitem[Liu et al.(2007)]{Liu2007} Liu, Q.~Z., van Paradijs, J., \& van den Heuvel, E.~P.~J.\ 2007, \aap, 469, 807

\bibitem[Liu \& Lyu(2021)]{LL2021} Liu, T. \& Lyu, K.-F.\ 2021, arXiv:2107.09971

\bibitem[Liu et al.(2021)]{Liu2021} Liu, T., Wei, Y.-F., Xue, L., et al.\ 2021, \apj, 908, 106

\bibitem[Lu et al.(2021)]{Lu2021} Lu, W., Beniamini, P., \& Bonnerot, C.\ 2021, \mnras, 500, 1817

\bibitem[Luo et al.(2016)]{Luo2016} Luo, J., Chen, L.-S., Duan, H.-Z., et al.\ 2016, Classical and Quantum Gravity, 33, 035010

\bibitem[MacLeod \& Ramirez-Ruiz(2015)]{MacLeod2015} MacLeod, M. \& Ramirez-Ruiz, E.\ 2015, \apjl, 798, L19

\bibitem[Marchant et al.(2021)]{Marchant2021} Marchant, P., Pappas, K.~M.~W., Gallegos-Garcia, M., et al.\ 2021, \aap, 650, A107

\bibitem[Mahy et al.(2022)]{Mahy2022} Mahy, L., Sana, H., Shenar, T., et al.\ 2022, \aap, 664, A159

\bibitem[Mandel(2016)]{Mandel2016} Mandel, I.\ 2016, \mnras, 456, 578

\bibitem[Mandel \& M{\"u}ller(2020)]{Mandel2020} Mandel, I. \& M{\"u}ller, B.\ 2020, \mnras, 499, 3214

\bibitem[Mandel et al.(2021)]{Mandel2021} Mandel, I., M{\"u}ller, B., Riley, J., et al.\ 2021, \mnras, 500, 1380

\bibitem[Mao et al.(2002)]{Mao2002} Mao, S., Smith, M.~C., Wo{\'z}niak, P., et al.\ 2002, \mnras, 329, 349

\bibitem[Mapelli \& Giacobbo(2018)]{Mapelli2018} Mapelli, M. \& Giacobbo, N.\ 2018, \mnras, 479, 4391

\bibitem[Margalit \& Metzger(2017)]{Margalit2017} Margalit, B. \& Metzger, B.~D.\ 2017, \apjl, 850, L19

\bibitem[Markert et al.(1973)]{Markert1973} Markert, T.~H., Canizares, C.~R., Clark, G.~W., et al.\ 1973, \apjl, 184, L67

\bibitem[Mashian \& Loeb(2017)]{Mashian2017} Mashian, N. \& Loeb, A.\ 2017, \mnras, 470, 2611

\bibitem[Michaely \& Perets(2016)]{Michaely2016} Michaely, E. \& Perets, H.~B.\ 2016, \mnras, 458, 4188

\bibitem[Moe \& Di Stefano(2017)]{Moe2017} Moe, M. \& Di Stefano, R.\ 2017, \apjs, 230, 15

\bibitem[Most et al.(2020)]{Most2020} Most, E.~R., Papenfort, L.~J., Weih, L.~R., et al.\ 2020, \mnras, 499, L82

\bibitem[Mr{\'o}z \& Wyrzykowski(2021)]{mw21} Mr{\'o}z, P. \& Wyrzykowski, {\L}.\ 2021, \actaa, 71, 89

\bibitem[Mr{\'o}z et al.(2021)]{Mroz2021} Mr{\'o}z, P., Udalski, A., Wyrzykowski, L., et al.\ 2021, arXiv:2107.13697

\bibitem[Mr{\'o}z et al.(2022)]{Mroz2022} Mr{\'o}z, P., Udalski, A., \& Gould, A.\ 2022, arXiv:2207.10729

\bibitem[Munar-Adrover et al.(2016)]{MA2016} Munar-Adrover, P., Sabatini, S., Piano, G., et al.\ 2016, \apj, 829, 101

\bibitem[Mu{\~n}oz-Darias et al.(2008)]{MD2008} Mu{\~n}oz-Darias, T., Casares, J., \& Mart{\'\i}nez-Pais, I.~G.\ 2008, \mnras, 385, 2205

\bibitem[Nan et al.(2011)]{Nan2011} Nan, R., Li, D., Jin, C., et al.\ 2011, International Journal of Modern Physics D, 20, 989

\bibitem[Naoz et al.(2016)]{Naoz2016} Naoz, S., Fragos, T., Geller, A., et al.\ 2016, \apjl, 822, L24

\bibitem[Nelemans et al.(2001)]{Nelemans2001} Nelemans, G., Yungelson, L.~R., \& Portegies Zwart, S.~F.\ 2001, \aap, 375, 890

\bibitem[Nelson \& Eggleton(2001)]{Nelson2001} Nelson, C.~A. \& Eggleton, P.~P.\ 2001, \apj, 552, 664

\bibitem[Neijssel et al.(2019)]{Neijssel2019} Neijssel, C.~J., Vigna-G{\'o}mez, A., Stevenson, S., et al.\ 2019, \mnras, 490, 3740

\bibitem[Neo et al.(1977)]{Neo1977} Neo, S., Miyaji, S., Nomoto, K., et al.\ 1977, \pasj, 29, 249

\bibitem[Nitz et al.(2021)]{Nitz2021} Nitz, A.~H., Kumar, S., Wang, Y.-F., et al.\ 2021, arXiv:2112.06878

\bibitem[Olejak et al.(2020)]{Olejak2020} Olejak, A., Belczynski, K., Bulik, T., et al.\ 2020, \aap, 638, A94

\bibitem[Olejak et al.(2021)]{Olejak2021} Olejak, A., Belczynski, K., \& Ivanova, N.\ 2021, \aap, 651, A100

\bibitem[Olejak et al.(2022)]{Olejak2022}  Olejak, A., Fryer, C.~L., Belczynski, K., et al.\ 2022, \mnras, 516, 2252

\bibitem[Orosz et al.(1998)]{Orosz1998} Orosz, J.~A., Jain, R.~K., Bailyn, C.~D., et al.\ 1998, \apj, 499, 375

\bibitem[Orosz et al.(2011)]{Orosz2011} Orosz, J.~A., McClintock, J.~E., Aufdenberg, J.~P., et al.\ 2011, \apj, 742, 84

\bibitem[{\"O}zel et al.(2010)]{ozel2010} {\"O}zel, F., Psaltis, D., Narayan, R., et al.\ 2010, \apj, 725, 1918

\bibitem[{\"O}zel \& Freire(2016)]{ozel2016} {\"O}zel, F. \& Freire, P.\ 2016, \araa, 54, 401

\bibitem[Packet(1981)]{Packet1981} Packet, W.\ 1981, \aap, 102, 17

\bibitem[Paczynski(1976)]{Paczynski1976} Paczynski, B.\ 1976, Structure and Evolution of Close Binary Systems, 73, 75

\bibitem[Paczynski(1986)]{Paczynski1986} Paczynski, B.\ 1986, \apj, 301, 503

\bibitem[Paczynski(1996)]{Paczynski1996} Paczynski, B.\ 1996, \araa, 34, 419

\bibitem[Pavlovskii \& Ivanova(2015)]{Pavlovskii2015} Pavlovskii, K. \& Ivanova, N.\ 2015, \mnras, 449, 4415

\bibitem[Pavlovskii et al.(2017)]{Pavlovskii2017} Pavlovskii, K., Ivanova, N., Belczynski, K., et al.\ 2017, \mnras, 465, 2092

\bibitem[Petrovic et al.(2005)]{Petrovic2005} Petrovic, J., Langer, N., \& van der Hucht, K.~A.\ 2005, \aap, 435, 1013


\bibitem[Podsiadlowski et al.(1992)]{Podsiadlowski1992} Podsiadlowski, P., Joss, P.~C., \& Hsu, J.~J.~L.\ 1992, \apj, 391, 246

\bibitem[Podsiadlowski et al.(1995)]{Podsiadlowski1995} Podsiadlowski, P., Cannon, R.~C., \& Rees, M.~J.\ 1995, \mnras, 274, 485

\bibitem[Podsiadlowski et al.(2003)]{Podsiadlowski2003} Podsiadlowski, P., Rappaport, S., \& Han, Z.\ 2003, \mnras, 341, 385

\bibitem[Podsiadlowski et al.(2010)]{Podsiadlowski2010} Podsiadlowski, P., Ivanova, N., Justham, S., et al.\ 2010, \mnras, 406, 840

\bibitem[Portegies Zwart et al.(1997)]{pz1997} Portegies Zwart, S.~F., Verbunt, F., \& Ergma, E.\ 1997, \aap, 321, 207

\bibitem[Porter \& Rivinius(2003)]{Porter2003} Porter, J.~M. \& Rivinius, T.\ 2003, \pasp, 115, 1153

\bibitem[Pylyser \& Savonije(1988)]{Pylyser1988} Pylyser, E. \& Savonije, G.~J.\ 1988, \aap, 191, 57

\bibitem[Pylyser \& Savonije(1989)]{Pylyser1989} Pylyser, E.~H.~P. \& Savonije, G.~J.\ 1989, \aap, 208, 52

\bibitem[Qin et al.(2018)]{Qin2018} Qin, Y., Fragos, T., Meynet, G., et al.\ 2018, \aap, 616, A28

\bibitem[Qin et al.(2019)]{Qin2019} Qin, Y., Marchant, P., Fragos, T., et al.\ 2019, \apjl, 870, L18

\bibitem[Qu \& Liu(2022)]{Qu2022} Qu, H.-M. \& Liu, T.\ 2022, \apj, 929, 83

\bibitem[Raithel et al.(2018)]{Raithel2018} Raithel, C.~A., Sukhbold, T., \& {\"O}zel, F.\ 2018, \apj, 856, 35

\bibitem[Remillard \& McClintock(2006)]{Remillard2006} Remillard, R.~A. \& McClintock, J.~E.\ 2006, \araa, 44, 49

\bibitem[Repetto et al.(2017)]{Repetto2017} Repetto, S., Igoshev, A.~P., \& Nelemans, G.\ 2017, \mnras, 467, 298

\bibitem[Reynolds et al.(2007)]{Reynolds2007} Reynolds, M.~T., Callanan, P.~J., \& Filippenko, A.~V.\ 2007, \mnras, 374, 657

\bibitem[Rezzolla et al.(2018)]{Rezzolla2018} Rezzolla, L., Most, E.~R., \& Weih, L.~R.\ 2018, \apjl, 852, L25

\bibitem[Rib{\'o} et al.(2017)]{Ribo2017} Rib{\'o}, M., Munar-Adrover, P., Paredes, J.~M., et al.\ 2017, \apjl, 835, L33

\bibitem[Rivinius et al.(2022)]{Rivinius2022} Rivinius, T., Klement, R., Chojnowski, S.~D., et al.\ 2022, arXiv:2208.12315

\bibitem[Romani et al.(2022)]{Romani2022} Romani, R.~W., Kandel, D., Filippenko, A.~V., et al.\ 2022, \apjl, 934, L17

\bibitem[Rowan et al.(2021)]{Rowan2021} Rowan, D.~M., Stanek, K.~Z., Jayasinghe, T., et al.\ 2021, \mnras, 507, 104

\bibitem[Ruan et al.(2020)]{Ruan2020} Ruan, W.-H., Guo, Z.-K., Cai, R.-G., et al.\ 2020, International Journal of Modern Physics A, 35, 2050075

\bibitem[Ruiz et al.(2018)]{Ruiz2018} Ruiz, M., Shapiro, S.~L., \& Tsokaros, A.\ 2018, \prd, 97, 021501

\bibitem[Sahu et al.(2022)]{Sahu2022} Sahu, K.~C., Anderson, J., Casertano, S., et al.\ 2022, \apj, 933, 83

\bibitem[Salpeter(1955)]{Salpeter1955} Salpeter, E.~E.\ 1955, \apj, 121, 161

\bibitem[Sana et al.(2012)]{Sana2012} Sana, H., de Mink, S.~E., de Koter, A., et al.\ 2012, Science, 337, 444

\bibitem[Saracino et al.(2022)]{Saracino2022} Saracino, S., Kamann, S., Guarcello, M.~G., et al.\ 2022, \mnras, 511, 2914

\bibitem[Schootemeijer et al.(2018)]{Schootemeijer2018} Schootemeijer, A., G{\"o}tberg, Y., de Mink, S.~E., et al.\ 2018, \aap, 615, A30

\bibitem[Seymour \& Yagi(2018)]{Seymour2018} Seymour, B. \& Yagi, K.\ 2018, \prd, 98, 124007

\bibitem[Seymour \& Yagi(2020)]{Seymour2020} Seymour, B.~C. \& Yagi, K.\ 2020, \prd, 102, 104003

\bibitem[Shahaf et al.(2022)]{Shahaf2022} Shahaf, S., Bashi, D., Mazeh, T., et al.\ 2022, arXiv:2209.00828

\bibitem[Shao et al.(2020)]{Shao2020} Shao, D.-S., Tang, S.-P., Jiang, J.-L., et al.\ 2020, \prd, 102, 063006

\bibitem[Shao \& Li(2014)]{sl2014} Shao, Y. \& Li, X.-D.\ 2014, \apj, 796, 37

\bibitem[Shao \& Li(2016)]{sl2016} Shao, Y. \& Li, X.-D.\ 2016, \apj, 833, 108

\bibitem[Shao \& Li(2018a)]{sl2018a} Shao, Y. \& Li, X.-D.\ 2018a, \mnras, 477, L128

\bibitem[Shao \& Li(2018b)]{sl2018b} Shao, Y. \& Li, X.-D.\ 2018b, \apj, 867, 124

\bibitem[Shao \& Li(2019)]{sl2019} Shao, Y. \& Li, X.-D.\ 2019, \apj, 885, 151

\bibitem[Shao \& Li(2020)]{sl2020} Shao, Y. \& Li, X.-D.\ 2020, \apj, 898, 143

\bibitem[Shao \& Li(2021a)]{sl2021a} Shao, Y. \& Li, X.-D.\ 2021a, \apj, 908, 67

\bibitem[Shao \& Li(2021b)]{sl2021b} Shao, Y. \& Li, X.-D.\ 2021b, \apj, 920, 81

\bibitem[Shao \& Li(2022)]{sl2022} Shao, Y. \& Li, X.-D.\ 2022, \apj, 930, 26

\bibitem[Shenar et al.(2020)]{Shenar2020} Shenar, T., Bodensteiner, J., Abdul-Masih, M., et al.\ 2020, \aap, 639, L6

\bibitem[Shenar et al.(2022a)]{Shenar2022a} Shenar, T., Sana, H., Mahy, L., et al.\ 2022a, arXiv:2207.07674

\bibitem[Shenar et al.(2022b)]{Shenar2022b} Shenar, T., Sana, H., Mahy, L., et al.\ 2022b, Nature Astronomy, 6, 1085

\bibitem[Shibata et al.(2019)]{Shibata2019} Shibata, M., Zhou, E., Kiuchi, K., et al.\ 2019, \prd, 100, 023015

\bibitem[Shikauchi et al.(2020)]{Shikauchi2020} Shikauchi, M., Kumamoto, J., Tanikawa, A., et al.\ 2020, \pasj, 72, 45

\bibitem[Shikauchi et al.(2022)]{Shikauchi2022} Shikauchi, M., Tanikawa, A., \& Kawanaka, N.\ 2022, \apj, 928, 13

\bibitem[Siegel et al.(2022)]{Siegel2022} Siegel, J.~C., Kiato, I., Kalogera, V., et al.\ 2022, arXiv:2209.06844


\bibitem[Soberman et al.(1997)]{Soberman1997} Soberman, G.~E., Phinney, E.~S., \& van den Heuvel, E.~P.~J.\ 1997, \aap, 327, 620

\bibitem[Soker \& Gilkis(2018)]{Soker2018} Soker, N. \& Gilkis, A.\ 2018, \mnras, 475, 1198

\bibitem[Soker et al.(2019)]{Soker2019} Soker, N., Grichener, A., \& Gilkis, A.\ 2019, \mnras, 484, 4972

\bibitem[Stancliffe \& Eldridge(2009)]{Stancliffe2009} Stancliffe, R.~J. \& Eldridge, J.~J.\ 2009, \mnras, 396, 1699

\bibitem[Su et al.(2021)]{Su2021} Su, B., Xianyu, Z.-Z., \& Zhang, X.\ 2021, \apj, 923, 114

\bibitem[Sukhbold et al.(2016)]{Sukhbold2016} Sukhbold, T., Ertl, T., Woosley, S.~E., et al.\ 2016, \apj, 821, 38

\bibitem[Tanikawa et al.(2022)]{Tanikawa2022} Tanikawa, A., Hattori, K., Kawanaka, N., et al.\ 2022, arXiv:2209.05632

\bibitem[Tauris et al.(2017)]{Tauris2017} Tauris, T.~M., Kramer, M., Freire, P.~C.~C., et al.\ 2017, \apj, 846, 170

\bibitem[Tauris(2018)]{Tauris2018} Tauris, T.~M.\ 2018, \prl, 121, 131105

\bibitem[Tauris \& van den Heuvel(2023)]{tv2023} Tauris, T. M., \& van den Heuvel, E. P. J. 2023, Physics of Binary Star Evolution (Princeton University Press)

\bibitem[The LIGO Scientific Collaboration et al.(2021)]{lvk2021} The LIGO Scientific Collaboration, the Virgo Collaboration, the KAGRA Collaboration, et al.\ 2021, arXiv:2111.03634

\bibitem[Thompson et al.(2019)]{Thompson2019} Thompson, T.~A., Kochanek, C.~S., Stanek, K.~Z., et al.\ 2019, Science, 366, 637

\bibitem[Timmes et al.(1996)]{Timmes1996} Timmes, F.~X., Woosley, S.~E., \& Weaver, T.~A.\ 1996, \apj, 457, 834

\bibitem[Tong et al.(2022)]{Tong2022} Tong, X., Wang, Y., \& Zhu, H.-Y.\ 2022, \apj, 924, 99

\bibitem[Trimble \& Thorne(1969)]{Trimble1969} Trimble, V.~L. \& Thorne, K.~S.\ 1969, \apj, 156, 1013

\bibitem[Tsokaros et al.(2020)]{Tsokaros2020} Tsokaros, A., Ruiz, M., \& Shapiro, S.~L.\ 2020, \apj, 905, 48

\bibitem[Tutukov \& Yungelson(1993)]{Tutukov1993} Tutukov, A.~V. \& Yungelson, L.~R.\ 1993, \mnras, 260, 675

\bibitem[Ugliano et al.(2012)]{Ugliano2012} Ugliano, M., Janka, H.-T., Marek, A., et al.\ 2012, \apj, 757, 69

\bibitem[Vanbeveren et al.(1979)]{Vanbeveren1979} Vanbeveren, D., de Gr{\`e}ve, J.~P., van Dessel, E.~L., et al.\ 1979, \aap, 73, 19

\bibitem[van den Heuvel(1975)]{vdh1975} van den Heuvel, E.~P.~J.\ 1975, \apjl, 198, L109

\bibitem[van den Heuvel(2009)]{vdh2009} van den Heuvel, E.~P.~J.\ 2009, Physics of Relativistic Objects in Compact Binaries: From Birth to Coalescence, 359, 125

\bibitem[van den Heuvel et al.(2017)]{vdh2017} van den Heuvel, E.~P.~J., Portegies Zwart, S.~F., \& de Mink, S.~E.\ 2017, \mnras, 471, 4256

\bibitem[van den Heuvel \& Tauris(2020)]{vt2020} van den Heuvel, E.~P.~J. \& Tauris, T.~M.\ 2020, Science, 368, eaba3282

\bibitem[van Kerkwijk et al.(1992)]{vanKerkwijk1992} van Kerkwijk, M.~H., Charles, P.~A., Geballe, T.~R., et al.\ 1992, \nat, 355, 703

\bibitem[van Paradijs(1996)]{vp1996} van Paradijs, J.\ 1996, \apjl, 464, L139

\bibitem[van Son et al.(2020)]{vs2020} van Son, L.~A.~C., De Mink, S.~E., Broekgaarden, F.~S., et al.\ 2020, \apj, 897, 100

\bibitem[van Son et al.(2022)]{vs2022} van Son, L.~A.~C., de Mink, S.~E., Callister, T., et al.\ 2022, \apj, 931, 17

\bibitem[Vietri \& Stella(1999)]{Vietri1999} Vietri, M. \& Stella, L.\ 1999, \apjl, 527, L43

\bibitem[Vigna-G{\'o}mez \& Ramirez-Ruiz(2022)]{vg2022} Vigna-G{\'o}mez, A. \& Ramirez-Ruiz, E.\ 2022, arXiv:2203.08478

\bibitem[Vinciguerra et al.(2020)]{Vinciguerra2020} Vinciguerra, S., Neijssel, C.~J., Vigna-G{\'o}mez, A., et al.\ 2020, \mnras, 498, 4705

\bibitem[Voss \& Tauris(2003)]{Voss2003} Voss, R. \& Tauris, T.~M.\ 2003, \mnras, 342, 1169

\bibitem[Wagg et al.(2021)]{Wagg2021} Wagg, T., Broekgaarden, F.~S., de Mink, S.~E., et al.\ 2021, arXiv:2111.13704

\bibitem[Wang et al.(2021)]{Wang2021} Wang, B., Chen, W.-C., Liu, D.-D., et al.\ 2021, \mnras, 506, 4654

\bibitem[Wang et al.(2016)]{Wang2016} Wang, C., Jia, K., \& Li, X.-D.\ 2016, \mnras, 457, 1015

\bibitem[Webb et al.(2000)]{Webb2000} Webb, N.~A., Naylor, T., Ioannou, Z., et al.\ 2000, \mnras, 317, 528

\bibitem[Webbink(1984)]{Webbink1984} Webbink, R.~F.\ 1984, \apj, 277, 355

\bibitem[Wellstein et al.(2001)]{Wellstein2001} Wellstein, S., Langer, N., \& Braun, H.\ 2001, \aap, 369, 939

\bibitem[Wiktorowicz et al.(2019)]{Wiktorowicz2019} Wiktorowicz, G., Wyrzykowski, {\L}., Chruslinska, M., et al.\ 2019, \apj, 885, 1

\bibitem[Wiktorowicz et al.(2020)]{Wiktorowicz2020} Wiktorowicz, G., Lu, Y., Wyrzykowski, {\L}., et al.\ 2020, \apj, 905, 134

\bibitem[Williams et al.(2010)]{Williams2010} Williams, S.~J., Gies, D.~R., Matson, R.~A., et al.\ 2010, \apjl, 723, L93

\bibitem[Woosley \& Weaver(1995)]{Woosley1995} Woosley, S.~E. \& Weaver, T.~A.\ 1995, \apjs, 101, 181

\bibitem[Wyrzykowski et al.(2016)]{Wyrzykowski2016} Wyrzykowski, {\L}., Kostrzewa-Rutkowska, Z., Skowron, J., et al.\ 2016, \mnras, 458, 3012

\bibitem[Wyrzykowski \& Mandel(2020)]{Wyrzykowski2020} Wyrzykowski, {\L}. \& Mandel, I.\ 2020, \aap, 636, A20

\bibitem[Xu \& Li(2018)]{Xu2018} Xu, X.-T. \& Li, X.-D.\ 2018, \apj, 859, 46

\bibitem[Yalinewich et al.(2018)]{Yalinewich2018} Yalinewich, A., Beniamini, P., Hotokezaka, K., et al.\ 2018, \mnras, 481, 930

\bibitem[Yamaguchi et al.(2018)]{Yamaguchi2018} Yamaguchi, M.~S., Kawanaka, N., Bulik, T., et al.\ 2018, \apj, 861, 21

\bibitem[Yang et al.(2020)]{Yang2020} Yang, Y., Gayathri, V., Bartos, I., et al.\ 2020, \apjl, 901, L34

\bibitem[Yi et al.(2019)]{Yi2019} Yi, T., Sun, M., \& Gu, W.-M.\ 2019, \apj, 886, 97

\bibitem[Yungelson \& Lasota(2008)]{Yungelson2008} Yungelson, L.~R. \& Lasota, J.-P.\ 2008, \aap, 488, 257

\bibitem[Zdziarski et al.(2013)]{Zdziarski2013} Zdziarski, A.~A., Mikolajewska, J., \& Belczynski, K.\ 2013, \mnras, 429, L104

\bibitem[Zevin et al.(2020)]{Zevin2020} Zevin, M., Spera, M., Berry, C.~P.~L., et al.\ 2020, \apjl, 899, L1

\bibitem[Zhang \& Li(2020)]{Zhang2020} Zhang, N.-B. \& Li, B.-A.\ 2020, \apj, 902, 38

\bibitem[Zheng et al.(2019)]{Zheng2019} Zheng, L.-L., Gu, W.-M., Yi, T., et al.\ 2019, \aj, 158, 179

\bibitem[Zhou et al.(2021)]{Zhou2021} Zhou, X., Li, A., \& Li, B.-A.\ 2021, \apj, 910, 62

\bibitem[Zhu et al.(2022)]{Zhu2022} Zhu, J.-P., Wu, S., Qin, Y., et al.\ 2022, \apj, 928, 167

\bibitem[Zuo et al.(2014)]{Zuo2014} Zuo, Z.-Y., Li, X.-D., \& Gu, Q.-S.\ 2014, \mnras, 437, 1187

\end{thebibliography}

\label{lastpage}

\end{document}